Preprint:

# A power series expansion of Teukolsky linearised gravitational waves.

Leo Brewin

School of Mathematics
Monash University, 3800
Australia

**Abstract**

A power series, suitable for use in a numerical relativity code, will be presented for the time symmetric Teukolsky linearised gravitational waves.

## 1 The Teukolsky metric

Any numerical relativity code worth its salt must demonstrate long term stability for initial data based on weak perturbations around flat spacetimes. A popular example of such spacetimes are the Teukolsky family of linearised gravitational waves [1]. One member of that family, and the main focus of this note, is described by the metric

$$\begin{aligned} ds^2 = &- dt^2 + dr^2 + r^2 d\Omega^2 \\ &+ \left(2 - 3\sin^2\theta\right) A(t,r) dr^2 \\ &- \left(A(t,r) - 3(\sin^2\theta) C(t,r)\right) r^2 d\theta^2 \\ &- \left(A(t,r) + 3(\sin^2\theta)\left(C(t,r) - A(t,r)\right)\right) r^2 \sin^2\theta d\phi^2 \\ &- 6r\left(\sin\theta\cos\theta\right) B(t,r) dr d\theta \end{aligned} \tag{1.1}$$

where

$$A(t,r) = \frac{3}{r^5}\left(r^2 F^{(2)} - 3rF^{(1)} + 3F\right) \tag{1.2}$$

$$B(t,r) = \frac{-1}{r^5}\left(-r^3 F^{(3)} + 3r^2 F^{(2)} - 6rF^{(1)} + 6F\right) \tag{1.3}$$

$$C(t,r) = \frac{1}{4r^5}\left(r^4 F^{(4)} - 2r^3 F^{(3)} + 9r^2 F^{(2)} - 21rF^{(1)} + 21F\right) \tag{1.4}$$

$$F^{(n)} = \frac{1}{2}\left(\frac{d^n Q(t+r)}{dr^n} - \frac{d^n Q(t-r)}{dr^n}\right) \tag{1.5}$$

and where $Q(x)$ is an arbitrary function of $x$. Note that this form of the metric differs slightly from that given by Teukolsky. Here the function $F$ has been expressed as an explicit combination of ingoing and outgoing waves (thus ensuring time symmetric initial data). Note also that the derivatives of $F$ are taken with respect to $r$ rather than $x$ as used by Teukolsky. Consequently, the signs of the odd-derivatives of $F$ in the expressions for $A$, $B$ and $C$ have been flipped (see Teukolsky's comment, last sentence on page 746).

Baumgarte and Shapiro [2] chose

$$Q(x) = axe^{-x^2} \tag{1.6}$$

where $a$ is a wave parameter (typically in the range $0 < a < 1$). Note also that this metric is not an exact solution of the vacuum Einstein equations but rather they yield $G_{ab} = \mathcal{O}(a^2)$ for $0 < a < 1$ where $G_{ab}$ are the components of the Einstein tensor.

There are two immediate problems in using the above form of the metric for supplying numerical initial data. First, the underlying spherical polar coordinates are singular at $r = 0$ and second, almost all numerical relativity codes are based on Cartesian like coordinates. One solution to this pair of problems is to sample the metric at points away from $r = 0$ and to then interpolate that data to a Cartesian grid. Though this can be made to work for the metric components, the interpolation will fail for other quantities like the connection or the Riemann curvatures. A better approach is to express the metric in terms of the Cartesian coordinates $(t, x, y, z)$. This is a standard (though tedious) computational task. The result is

$$\begin{aligned}
ds^2 = &- dt^2 + dx^2 + dy^2 + dz^2 \\
&+ \frac{a}{r^4}\left((3x^2z^2 + 3y^2r^2 - r^4)A(t,r) - 6x^2z^2B(t,r) + 3(x^2z^2 - y^2r^2)C(t,r)\right)dx^2 \\
&+ \frac{a}{r^4}\left((3y^2z^2 + 3x^2r^2 - r^4)A(t,r) - 6y^2z^2B(t,r) + 3(y^2z^2 - x^2r^2)C(t,r)\right)dy^2 \\
&+ \frac{a}{r^4}\left((3z^4 - r^4)A(t,r) + 6z^2(x^2+y^2)B(t,r) + 3(x^2+y^2)^2C(t,r)\right)dz^2 \\
&- 6\frac{a}{r^4}\left(xy(x^2+y^2)A(t,r) + 2xyz^2B(t,r) - xy(r^2+z^2)C(t,r)\right)dxdy \\
&+ 6\frac{a}{r^4}\left(xz^3A(t,r) + xz(x^2+y^2-z^2)B(t,r) - xz(x^2+y^2)C(t,r)\right)dxdz \\
&+ 6\frac{a}{r^4}\left(yz^3A(t,r) + yz(x^2+y^2-z^2)B(t,r) - yz(x^2+y^2)C(t,r)\right)dydz
\end{aligned} \tag{1.7}$$

In this form it is clear that there may be numerical problems at $r = 0$. The functions $A, B$ and $C$ each carry a $1/r^5$ factor (as per equations (1.2–1.4)) while the conversion to Cartesian coordinates has introduced a further factor of $1/r^4$. Thus near the origin there is a possible $1/r^9$ behaviour in the metric components. However, the underlying physics (of weak gravitational waves) suggests that the metric should be non-singular at $r = 0$ and thus the singular terms should somehow cancel out.

The remainder of this note will develop a power series expansion around $r = 0$ to explicitly show that this is the case. That same series can then be used to supply initial data on and near $r = 0$ without any numerical concerns.

# 2 Series expansion of the metric.

The objective here is to use a power series around $r = 0$ to show explicitly that the $A, B, C$ and $g_{ab}$ are all well behaved near $r = 0$.

Notice that the $g_{ab}$ depend linearly on $A, B$ and $C$ which in turn depend linearly on $F$. Thus a power series for the $g_{ab}$ can be inferred from a power series for $F$.

Begin by writing $F$ in the form

$$F(t,r) = a\left((t+r)e^{-2tr} - (t-r)e^{2tr}\right)e^{-t^2-r^2} \tag{2.1}$$

and then use a standard power series for the $e^{-2tr}$ and $e^{2tr}$ terms to obtain

$$e^{t^2+r^2}F = a(2-4t^2)r + a2^3t^2\left(\frac{1}{2!} - \frac{2t^2}{3!}\right)r^3 + a2^5t^4\left(\frac{1}{4!} - \frac{2t^2}{5!}\right)r^5 + \dots \tag{2.2}$$

$$= a\sum_{n=0}^{\infty} f_n r^{2n+1} \tag{2.3}$$

with

$$f_n = 2(2n+1-2t^2)\frac{(2t)^{2n}}{(2n+1)!} \tag{2.4}$$

The remaining steps are straightforward – simply substitute the series expansion for $F$ (2.2,2.4) into the equations for $A, B$ and $C$ (1.2–1.4) and finally the metric (1.7). The results for the three functions $A, B$ and $C$ are

$$A = ae^{-t^2-r^2}\sum_{n=0}^{\infty} f_n a_n \tag{2.5}$$

$$B = ae^{-t^2-r^2}\sum_{n=0}^{\infty} f_n b_n \tag{2.6}$$

$$C = ae^{-t^2-r^2}\sum_{n=0}^{\infty} f_n c_n \tag{2.7}$$

where

$$\begin{aligned}
a_n &= 12nr^{2n-4}(n-1) - 24nr^{2n-2} + 12r^{2n}\\
b_n &= -24n^2r^{2n-2} + 4nr^{2n-4}(n-1)(2n-1) + 12r^{2n}(2n+1) - 8r^{2n+2}\\
c_n &= 4nr^{2n-4}(n-1)\left(n^2-n+1\right) - 8nr^{2n-2}\left(2n^2+1\right)\\
&\quad + 12r^{2n}\left(2n^2+2n+1\right) - 16r^{2n+2}(n+1) + 4r^{2n+4}
\end{aligned}$$

The full power series for the $g_{ab}$ is given by

$$g_{ab} = \eta_{ab} + ae^{-t^2-r^2}\sum_{n=0}^{\infty} f_n g_{ab}^{(n)} \tag{2.8}$$

where

$$\begin{aligned}
g_{xx}^{(n)} = &-12nr^{2n-8}x^2(n-3)(n-2)(n-1)\left(x^2+y^2\right)\\
&+ 12nr^{2n-6}(n-2)(n-1)\left(nx^2 - ny^2 + 4x^4 + 4x^2y^2 - 3x^2 - y^2\right)\\
&- 12nr^{2n-4}(n-1)\left(4nx^2 - 4ny^2 + 6x^4 + 6x^2y^2 - 8x^2 - 4y^2 + 1\right)\\
&+ 24nr^{2n-2}\left(3nx^2 - 3ny^2 + 2x^4 + 2x^2y^2 - 3x^2 - 3y^2 + 1\right)\\
&- 12r^{2n}\left(4nx^2 - 4ny^2 + x^4 + x^2y^2 - 4y^2 + 1\right) + 12r^{2n+2}(x-y)(x+y)
\end{aligned}$$

$$
\begin{aligned}
g_{yy}^{(n)} &= -12nr^{2n-8}y^2\,(n-3)\,(n-2)\,(n-1)\left(x^2+y^2\right)\\
&\quad+12nr^{2n-6}\,(n-2)\,(n-1)\left(-nx^2+ny^2+4x^2y^2-x^2+4y^4-3y^2\right)\\
&\quad-12nr^{2n-4}\,(n-1)\left(-4nx^2+4ny^2+6x^2y^2-4x^2+6y^4-8y^2+1\right)\\
&\quad+24nr^{2n-2}\left(-3nx^2+3ny^2+2x^2y^2-3x^2+2y^4-3y^2+1\right)\\
&\quad-12r^{2n}\left(-4nx^2+4ny^2+x^2y^2-4x^2+y^4+1\right)-12r^{2n+2}\,(x-y)\,(x+y)
\end{aligned}
$$

$$
\begin{aligned}
g_{zz}^{(n)} &= 12nr^{2n-8}\,(n-3)\,(n-2)\,(n-1)\left(x^2+y^2\right)^2-48nr^{2n-6}\,(n-2)\,(n-1)\left(x^2+y^2\right)\left(x^2+y^2-1\right)\\
&\quad+24nr^{2n-4}\,(n-1)\left(3x^4+6x^2y^2-6x^2+3y^4-6y^2+1\right)\\
&\quad-48nr^{2n-2}\left(x^4+2x^2y^2-3x^2+y^4-3y^2+1\right)+12r^{2n}\left(x^4+2x^2y^2-4x^2+y^4-4y^2+2\right)
\end{aligned}
$$

$$
\begin{aligned}
g_{xy}^{(n)} &= -12nr^{2n-8}xy\,(n-3)\,(n-2)\,(n-1)\left(x^2+y^2\right)\\
&\quad+24nr^{2n-6}xy\,(n-2)\,(n-1)\left(n+2x^2+2y^2-1\right)-24nr^{2n-4}xy\,(n-1)\left(4n+3x^2+3y^2-2\right)\\
&\quad+48nr^{2n-2}xy\left(3n+x^2+y^2\right)-12r^{2n}xy\left(8n+x^2+y^2+4\right)+24r^{2n+2}xy
\end{aligned}
$$

$$
\begin{aligned}
g_{xz}^{(n)} &= -12nr^{2n-8}xz\,(n-3)\,(n-2)\,(n-1)\left(x^2+y^2\right)+24nr^{2n-6}xz\,(n-2)\,(n-1)\left(2x^2+2y^2-1\right)\\
&\quad-72nr^{2n-4}xz\,(n-1)\left(x^2+y^2-1\right)+24nr^{2n-2}xz\left(2x^2+2y^2-3\right)-12r^{2n}xz\left(x^2+y^2-2\right)
\end{aligned}
$$

$$
\begin{aligned}
g_{yz}^{(n)} &= -12nr^{2n-8}yz\,(n-3)\,(n-2)\,(n-1)\left(x^2+y^2\right)+24nr^{2n-6}yz\,(n-2)\,(n-1)\left(2x^2+2y^2-1\right)\\
&\quad-72nr^{2n-4}yz\,(n-1)\left(x^2+y^2-1\right)+24nr^{2n-2}yz\left(2x^2+2y^2-3\right)-12r^{2n}yz\left(x^2+y^2-2\right)
\end{aligned}
$$

Note that the above $g_{ab}^{(n)}$ are manifestly non-singular at $r=0$.

# 3 Convergence.

The results presented in the previous section will be of little use if the series do not converge in a suitable domain containing the origin $r=0$. It is actually quite easy to demonstrate, using a standard ratio test, that each of the above power series converges absolutely for all (finite) $x, y, z, t$. Consider, for example, the convergence of the series for $g_{xy}$. A straightforward computation shows, for any finite $x, y, z, t$, that

$$
\lim_{n\to\infty}\left|\frac{f_{n+1}g_{xy}^{(n+1)}}{f_n g_{xy}^{(n)}}\right| = \lim_{n\to\infty}\left|\frac{4t^2r^2}{2n+3}\right| = 0 \tag{3.1}
$$

Thus, as claimed, the power series for $g_{xy}$ is absolutely convergent. The same method can be applied to the remaining power series (with the same result).

# 4 Explicit form of the metric.

The power series would only be required when evaluating the metric components near the origin. Everywhere else, the metric components would be evaluated directly from (1.2–1.7). The resulting expression for the full metric is a dog's breakfast, however, some semblance of order can be had by writing the results as the sum of two terms, one for each of the exponentials, as follows

$$
g_{ab} = a\left(g_{ab}^{+}e^{-(r+t)^2}+g_{ab}^{-}e^{-(r-t)^2}\right) \tag{4.1}
$$

where

$$
\begin{aligned}
4r^9(g^+_{xx}-1) = {} & -48r^{11}y^2 - 240r^{10}ty^2 + 48r^9\left(-10t^2y^2 + x^2z^2 + 4y^2 - 1\right) \\
& + 48r^8t\left(-10t^2y^2 + 5x^2z^2 + 11y^2 - 3\right) + 48r^7t^2\left(-5t^2y^2 + 10x^2z^2 + 9y^2 - 3\right) \\
& + 24r^6t\left(-2t^4y^2 + 20t^2x^2z^2 + 2t^2y^2 - 2t^2 + 10x^2z^2 + 9y^2 - 3\right) \\
& + 12r^5t^2\left(20t^2x^2z^2 - 4t^2y^2 + 60x^2z^2 + 21y^2 - 6\right) \\
& + 6r^4t\left(8t^4x^2z^2 + 120t^2x^2z^2 + 6t^2y^2 + 60x^2z^2 + 21y^2 - 6\right) \\
& + 30r^3t^2\left(8t^2x^2z^2 + 30x^2z^2 + 3y^2\right) \\
& + 45r^2t\left(12t^2x^2z^2 + 10x^2z^2 + y^2\right) + 630rt^2x^2z^2 + 315tx^2z^2
\end{aligned}
$$

$$
\begin{aligned}
4r^9(g^+_{yy}-1) = {} & -48r^{11}x^2 - 240r^{10}tx^2 - 48r^9\left(10t^2x^2 - 4x^2 - y^2z^2 + 1\right) \\
& - 48r^8t\left(10t^2x^2 - 11x^2 - 5y^2z^2 + 3\right) - 48r^7t^2\left(5t^2x^2 - 9x^2 - 10y^2z^2 + 3\right) \\
& - 24r^6t\left(2t^4x^2 - 2t^2x^2 - 20t^2y^2z^2 + 2t^2 - 9x^2 - 10y^2z^2 + 3\right) \\
& - 12r^5t^2\left(4t^2x^2 - 20t^2y^2z^2 - 21x^2 - 60y^2z^2 + 6\right) \\
& + 6r^4t\left(8t^4y^2z^2 + 6t^2x^2 + 120t^2y^2z^2 + 21x^2 + 60y^2z^2 - 6\right) \\
& + 30r^3t^2\left(8t^2y^2z^2 + 3x^2 + 30y^2z^2\right) \\
& + 45r^2t\left(12t^2y^2z^2 + x^2 + 10y^2z^2\right) + 630rt^2y^2z^2 + 315ty^2z^2
\end{aligned}
$$

$$
\begin{aligned}
4r^9(g^+_{zz}-1) = {} & 48r^9\left(s^4 - 4s^2 + 2\right) + 48r^8t\left(5s^4 - 16s^2 + 6\right) + 96r^7t^2\left(5s^4 - 12s^2 + 3\right) \\
& + 48r^6t\left(10s^4t^2 + 5s^4 - 16s^2t^2 - 12s^2 + 2t^2 + 3\right) \\
& + 48r^5t^2\left(5s^4t^2 + 15s^4 - 4s^2t^2 - 24s^2 + 3\right) \\
& + 24r^4t\left(2s^4t^4 + 30s^4t^2 + 15s^4 - 24s^2t^2 - 24s^2 + 3\right) \\
& + 60r^3s^2t^2\left(4s^2t^2 + 15s^2 - 12\right) + 90r^2s^2t\left(6s^2t^2 + 5s^2 - 4\right) + 630rs^4t^2 + 315s^4t
\end{aligned}
$$

$$
\begin{aligned}
4r^9g^+_{xy} = {} & 96r^{11}xy + 480r^{10}txy - 48r^9xy\left(s^2 - 20t^2 + 4\right) - 48r^8txy\left(5s^2 - 20t^2 + 6\right) \\
& - 96r^7t^2xy\left(5s^2 - 5t^2 - 3\right) - 48r^6txy\left(10s^2t^2 + 5s^2 - 2t^4 - 14t^2 - 3\right) \\
& - 24r^5t^2xy\left(10s^2t^2 + 30s^2 - 12t^2 - 27\right) - 12r^4txy\left(4s^2t^4 + 60s^2t^2 + 30s^2 - 42t^2 - 27\right) \\
& - 60r^3t^2xy\left(4s^2t^2 + 15s^2 - 9\right) - 90r^2txy\left(6s^2t^2 + 5s^2 - 3\right) - 630rs^2t^2xy - 315s^2txy
\end{aligned}
$$

$$
\begin{aligned}
4r^9g^+_{xz} = {} & -48r^9xz\left(s^2 - 2\right) - 48r^8txz\left(5s^2 - 8\right) - 96r^7t^2xz\left(5s^2 - 6\right) \\
& - 48r^6txz\left(10s^2t^2 + 5s^2 - 8t^2 - 6\right) - 48r^5t^2xz\left(5s^2t^2 + 15s^2 - 2t^2 - 12\right) \\
& - 24r^4txz\left(2s^2t^4 + 30s^2t^2 + 15s^2 - 12t^2 - 12\right) - 60r^3t^2xz\left(4s^2t^2 + 15s^2 - 6\right) \\
& - 90r^2txz\left(6s^2t^2 + 5s^2 - 2\right) - 630rs^2t^2xz - 315s^2txz
\end{aligned}
$$

$$
\begin{aligned}
4r^9g^+_{yz} = {} & -48r^9yz\left(s^2 - 2\right) - 48r^8tyz\left(5s^2 - 8\right) - 96r^7t^2yz\left(5s^2 - 6\right) - 48r^6tyz\left(10s^2t^2 + 5s^2 - 8t^2\right. \\
& \left.- 6\right) - 48r^5t^2yz\left(5s^2t^2 + 15s^2 - 2t^2 - 12\right) - 24r^4tyz\left(2s^2t^4 + 30s^2t^2 + 15s^2 - 12t^2 - 12\right) \\
& - 60r^3t^2yz\left(4s^2t^2 + 15s^2 - 6\right) - 90r^2tyz\left(6s^2t^2 + 5s^2 - 2\right) - 630rs^2t^2yz - 315s^2tyz
\end{aligned}
$$

and where $s^2 = x^2+y^2$ and $r^2 = x^2+y^2+z^2$. The $g^-_{ab}$ can be obtained by the simple substitution $t \to -t$.

# References

[1] S. A. Teukolsky, Linearized quadrupole waves in general relativity and the motion of test particles, *Phys. Rev. D* **26** 745–750.

[2] T. W. Baumgarte and S. L. Shapiro, Numerical integration of Einstein's field equations, *Phys. Rev. D* **59** 024007, gr-qc/9810065.